\documentclass{IEEEcsmag}

\usepackage[colorlinks,urlcolor=blue,linkcolor=blue,citecolor=blue]{hyperref}
\expandafter\def\expandafter\UrlBreaks\expandafter{\UrlBreaks\do\/\do\*\do\-\do\~\do\'\do\"\do\-}
\usepackage{upmath,color}
\usepackage[super,comma]{natbib}

\usepackage{multicol}

\usepackage{etoolbox}

\jname{Computer}
\jtitle{Cybertrust}
\pubyear{2025}

\begin{document}

\sptitle{Column - Agentic AI in Cybersecurity}

\title{Agentic AI and the Cyber Arms Race}

\author{Sean Oesch}
\affil{Oak Ridge National Laboratory, Oak Ridge, TN, USA, oeschts@ornl.gov}

\author{Jack Hutchins}
\affil{Oak Ridge National Laboratory, hutchinsjr@ornl.gov}

\author{Phillipe Austria}
\affil{Oak Ridge National Laboratory, austriaps@ornl.gov}

\author{Amul Chaulagain}
\affil{Oak Ridge National Laboratory, chaulagaina@ornl.gov}

\markboth{Agentic AI and the Cyber Arms Race}{Cybertrust Column/Computer Magazine 2025}

\begin{abstract}
\looseness0-Agentic AI is shifting the cybersecurity landscape as attackers and defenders leverage AI agents to augment humans and automate common tasks. In this article, we examine the implications for cyber warfare and global politics as Agentic AI becomes more powerful and enables the broad proliferation of capabilities only available to the most well resourced actors today. 
\end{abstract}

\maketitle

\chapteri{I}n the early years of cybersecurity defenders utilized virus-specific signatures, honeypots, and heuristics. 
As attacks increased in volume and attackers became more sophisticated, moving towards polymorphic malware, packers, and novel evasion techniques, defenders looked to machine learning to provide scalability (quickly analyze large volumes of data and automate repetitive tasks), pattern recognition (detect common attack patterns), and novelty detection (recognize abnormal behaviors that may indicate malicious actors or insider threats).
Companies now use Large Language Models (LLMs) to provide analysts and reverse engineers with a rapid analysis of malicious code and best next steps when triaging alerts.
\footnote{For example, see \href{https://www.microsoft.com/en-us/security/business/ai-machine-learning/microsoft-security-copilot}{Microsoft Security Copilot}}
But the real paradigm shift in cybersecurity for both attackers and defenders is still on the horizon: agentic artificial intelligence (agentic AI).\footnote{This work will be published in Computer Magazine in the Cybertrust column by IEEE.}

Cyber warfare is inherently different from traditional kinetic warfare.\footnote{See article \href{https://www.yalejournal.org/publications/cult-of-the-cyber-offensive-misperceptions-of-the-cyber-offensedefense-balance}{Cult of the Cyber Offensive: Misperceptions of the Cyber Offense/Defense Balance}}
Kinetic warfare seeks to force an opponent to submit through physical violence. 
Cyber warfare seeks a strategic advantage through espionage, disruption, and degradation of information and operational systems.  
In cyber warfare, skill is the weapon and cyber weapons suffer from impermanence --- they only work well once, or until the threat is publicly acknowledged and relevant systems patched to prevent the threat vector being exploited again. 
As defenders become more sophisticated, the cost of developing an effective cyber weapon becomes prohibitively expensive, so that only nation state actors or their equivalents can afford to leverage the skills necessary to create them. 

But what if the skills needed to create cyber weapons become widely available through AI agents?
While a single agent capable of replacing a skilled human is likely still be a decade or more away\footnote{See article \href{https://ourworldindata.org/ai-timelines}{Article from Our World in Data}}, an agent composed of multiple hierarchical models with specific skill sets is immanent. 
Imagine a Centralized Reinforcement Learning Agent (CARL) in control of a suite of task-specific agents: a LREM (Large Reverse Engineering Model) trained to understand, produce, and manipulate binary code, a log agent trained to digest and make inferences from disparate log data, a networking agent capable of mapping and traversing networks, and a vulnerability finder agent able to analyze a system or service and identify effective Tactics, Techniques, and Procedures (TTPs).
Each of these task-specific agents may utilize a multi-agent solution as well and interact with existing libraries and tools. 
CARL is now capable of achieving complex tasks by delegating work to the appropriate task-specific agent, achieving behaviors that mimic those of a skilled human. 

Existing multi-agent orchestrations platforms such as \href{https://www.crewai.com/}{CrewAI} already allow agents to work together to achieve complex tasks. 
And several skilled cyber-specific agents already exist. 
Companies like XBOW and RunSybil\footnote{\href{https://www.runsybil.com/}{RunSybil Website}} use AI agents to automate pentesting, with XBOW able to find and exploit vulnerabilities in 75\% of web security benchmarks and discover novel vulnerabilities in web applications.~\footnote{\href{https://xbow.com/blog/xbow-scoold-vuln/}{How XBOW found a Scoold authentication bypass}}
And \href{https://www.dropzone.ai/}{Dropzone AI} uses autonomous agents to automate alert triage and other Tier 1 tasks in the Security Operations Center. 

As these cyber agents become more powerful and are combined into multi-agent solutions, they have the potential to fundamentally change the nature of and shift the balance of power in the cyber landscape. 
In Section~\ref{sec:cyber}, we explore the ways that agentic AI may change the symmetry between attackers and defenders based upon the expertise we have developed during our own research~\cite{oesch2024towards, oesch2024path}.
And in Section~\ref{sec:politics}, we discuss the potential implications of agentic AI in geopolitics through a comparison and contrast with the proliferation of nuclear weapons. 

\section{Implications for the Balance of Power in Cyber Warfare}\label{sec:cyber}
Contemporary cybersecurity follows a cyclical pattern where absolute prevention of attacks remains unfeasible; threat actors exploit vulnerabilities, defenders respond with containment measures and patches, and both parties engage in an ongoing process of adaptation and learning. 
This creates a co-evolutionary dynamic between attackers and defenders, each developing increasingly better methods in response to the other's capabilities~\cite{Hoffman2021}. 
We believe the introduction of sophisticated Agentic AI into the cyber domain is likely to both shatter and maintain this fundamental pattern. 
It will maintain these pattern because AI agents for offense and defense possess capabilities for adaptation and evolution in response to one another.
When AI attack agents grow in capability, defensive AI agents can adapt through retraining or dynamic adaptive capabilities. 

However, Agentic AI agents may also shatter the existing paradigm by empowering previously insignificant actors and further exposing entities without the resources to update their own defenses. 
The cost of maintaining a strong defensive security posture is inherently higher than conducting an attack, especially if that attack is automated, and many organizations already lack the ability to withstand today's threats. 
Organizations without the money to buy or knowledge to implement effective defenses may be overwhelmed in this new world of Agentic AI. 
Moreover, if AI agents gain the ability to create new offensive attacks in hours, minutes, or even seconds against complex defensive systems, it may not be possible for defensive agents to adapt quickly enough to counter the threats, or defenders may be required to adopt more aggressive strategies, such as adversarial AI (discussed below) or actively finding and compromising adversary infrastructure. 

\begin{figure}
    \centering
    \includegraphics[width=1\linewidth]{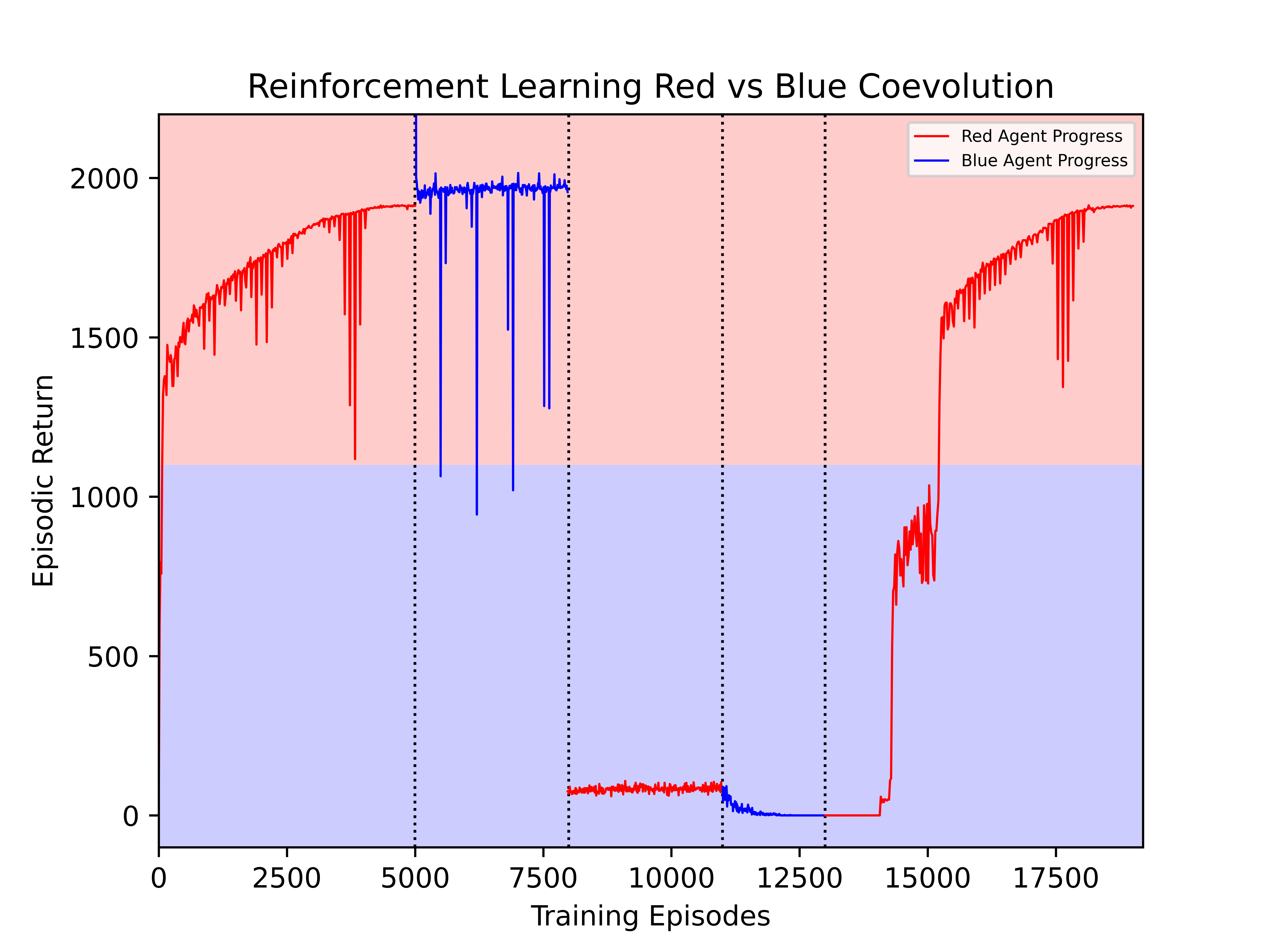}
    \caption{Graph showing AI Red/Blue Agent coevolution.
    A low episodic return indicates the blue agent is performing well, while a high episodic return indicates the red agent is performing well. 
    The vertical lines represent different training runs. 
    In the first run, the red agent is trained without a blue agent.
    Next, the blue agent is trained against this version of the red agent, and so on. 
    As can be seen, over multiple runs the agents can learn to adapt to changes in one another's abilities, effectively coevolving.
    Cyberwheel, the environment that generated this graph, is on \href{https://github.com/ORNL/cyberwheel}{ORNL's Github}. 
    }
    \label{fig:coevolution}
\end{figure}

Building on our prior research into adaptive cyber defense agents~\cite{oesch2024path, oesch2024towards}, we have begun to test the ability of attack and defense agents to adapt to one another in a co-evolutionary fashion. 
As can be seen in Figure~\ref{fig:coevolution}, offensive and defensive AI agents are capable of adapting to improvements in each others' capabilities simply by retraining after the opposing agent is updated. 
 
One key difference between traditional cybersecurity and Agentic AI is that AI is vulnerable to unique attacks against its robustness via adversarial AI. 
Parquini et al.~\cite{pasquini2024hacking} recently demonstrated this difference by tricking an LLM-based AI red agent. 
When the LLM-based agent attacked they were able to detect that it was an AI and they changed the system response to make the agent fail. 
This is only one example of the many ways that AI agents could be compromised and is a vulnerability whether they are being used for offense or defense. 
Countering adversarial AI and guaranteeing agent robustness will be essential to the future of Agentic AI in cyber. 

\section{Implications for Geopolitics}\label{sec:politics}
In addition to impacting the dynamic between offense and defense in cybersecurit, Agentic AI has huge potential to shift global power dynamics, much like the advent of nuclear weapons reshaped global power dynamics in the twentieth century. 
In the same way that the Manhattan Project heralded a new era of strategic deterrence, secrecy, and arms racing, the broader accessibility of AI-enabled capabilities could empower not only major powers but also smaller states and even non-state actors. 
Unlike nuclear technology, whose development hinged on large-scale, highly centralized projects under tight government control, key components of AI research, including open-source frameworks and off-the-shelf computing power, may be more diffusely available. 
This lowers the barrier to entry for countries eager to acquire a new form of deterrence or regional influence, much as smaller states historically leveraged disruptive military technologies to counterbalance conventional power asymmetries~\cite{schmid2018determinants}.

The result could be a two-tiered ecosystem, reminiscent of early atomic history, in which a handful of powerful actors retain access to the most cutting-edge models requiring immense data resources and sophisticated infrastructure, while mid-level states capitalize on “good-enough” AI to carry out disruptive operations. 
Much like the nuclear era ultimately extended beyond the initial superpowers despite attempts at containment \cite{burr2000whether}, it is unlikely that the most advanced AI capabilities will remain exclusive for long. 
Once foundational knowledge and baseline technologies become widely understood, ambitious states can funnel resources, both overtly and covertly, into domestic AI labs or forge alliances to acquire expertise. 
The historical diffusion of strategic technologies underscores how determined nations eventually bridge initial capability gaps, especially when regional ambitions or security dilemmas drive them forward \cite{schmid2018determinants}.

Smaller states that perfect even moderately sophisticated autonomous cyber operations could punch above their weight. 
A well-placed AI-driven intrusion could degrade vital infrastructure, extract sensitive data, or sow panic by manipulating information systems at key moments. 
This ability to project power in cyberspace mirrors how early nuclear programs granted regional players disproportionate diplomatic leverage \cite{gaddis1986long}. 
In a volatile geopolitical environment, autonomous cyber systems could serve as tools for smaller nations to disrupt or deter larger powers, achieving strategic objectives without the risks and costs associated with conventional military engagements.

From a geopolitical standpoint, this democratization effect is tempered by the tendency of leading powers to maintain an edge. 
Historically, strategic advantages in technology have often created a lag period during which dominant players consolidate their position before broader proliferation occurs. 
In cyberspace, however, the pace of innovation and the decentralized nature of AI development could compress this timeline, potentially resulting in rapid horizontal proliferation of mid-tier capabilities. 
Meanwhile, vertical proliferation (the constant refinement of top-tier AI systems by technologically advanced nations) will likely exacerbate global inequalities, creating an arms race dynamic that pushes smaller players to adopt asymmetrical tactics.

Unlike the nuclear age, where deterrence mechanisms and mutual assured destruction eventually stabilized conflict between superpowers, agentic AI introduces greater unpredictability. 
The opacity and speed of cyber operations complicate attribution, raising the likelihood of retaliatory actions based on suspicion rather than certainty. 
Without the transparency and verifiability that characterized Cold War-era arms treaties, the digital domain may foster an arms race with few constraints. 
Historical parallels to the “long peace” of the Cold War \cite{gaddis1986long} suggest that stability relied on a balance of power and clear communication—factors that are notably absent in cyberspace, where attacks often unfold covertly and instantaneously.

One critical divergence from past strategic technologies lies in the role of smaller states.
With AI, the cost of entry is lower, and the need for traditional industrial capabilities is reduced. 
This dynamic positions agentic AI as a potential equalizer, enabling less resourced nations to exert influence in their regions. 
Such states may use autonomous cyber tools to project power, deter aggression, or destabilize adversaries. 
This aligns with historical patterns of smaller nations leveraging disruptive technologies for strategic advantage, yet agentic AI’s versatility and speed could amplify these effects beyond what was possible with previous tools like ballistic missiles or drones.

Together, these developments suggest a future in which global order becomes more fluid and unpredictable than under the bipolar stability of the Cold War, or the unipolar stability that followed. 
Mutual assured destruction hinged on transparent demonstrations of nuclear potency, codified by treaties and shaped by crises that reinforced a balance of terror. 
In contrast, agentic AI developments and deployments evolve in secrecy, often without the oversight or accountability necessary to prevent escalation. 
History has shown that disruptive military technologies are rarely contained indefinitely \cite{burr2000whether}, and agentic AI appears poised to follow this trajectory. 
Whether its proliferation yields more frequent low-level cyber skirmishes or destabilizing conflicts among major powers remains uncertain. 
However, what is clear is that autonomous cyber capabilities will augment the arsenals of dominant players while empowering smaller and emerging states to assert themselves in ways that echo, and may also eclipse, the transformations wrought by nuclear weapons in the twentieth century.

\vspace*{-8pt}

\section{ACKNOWLEDGMENTS}
Notice: This manuscript has been authored by UT-Battelle, LLC under Contract No. DE-AC05-00OR22725 with the U.S. Department of Energy. The United States Government retains and the publisher, by accepting the article for publication, acknowledges that the United States Government retains a non-exclusive, paid-up, irrevocable, world-wide license to publish or reproduce the published form of this manuscript, or allow others to do so, for United States Government purposes. The Department of Energy will provide public access to these results of federally sponsored research in accordance with the DOE Public Access Plan (http://energy.gov/downloads/doe-public-access-plan).

This manuscript was prepared as part of the Emerging and Cyber Security Technologies initiative at Oak Ridge National Laboratory. 

\def\refname{REFERENCES}

\bibliographystyle{plainnat}
\bibliography{references}

\begin{thebibliography}{7}
\providecommand{\natexlab}[1]{#1}
\providecommand{\url}[1]{\texttt{#1}}
\expandafter\ifx\csname urlstyle\endcsname\relax
  \providecommand{\doi}[1]{doi: #1}\else
  \providecommand{\doi}{doi: \begingroup \urlstyle{rm}\Url}\fi

\bibitem[Burr and Richelson(2000)]{burr2000whether}
William Burr and Jeffrey~T Richelson.
\newblock Whether to" strangle the baby in the cradle": The united states and
  the chinese nuclear program, 1960-64.
\newblock \emph{International Security}, 25\penalty0 (3):\penalty0 54--99,
  2000.

\bibitem[Gaddis(1986)]{gaddis1986long}
John~Lewis Gaddis.
\newblock The long peace: Elements of stability in the postwar international
  system.
\newblock \emph{International security}, 10\penalty0 (4):\penalty0 99--142,
  1986.

\bibitem[Hoffman(2021)]{Hoffman2021}
Wyatt Hoffman.
\newblock Ai and the future of cyber competition.
\newblock In \emph{Issue Brief, Center for Security and Emerging Technology
  (CSET)}, 01 2021.
\newblock URL
  \url{https://cset.georgetown.edu/publication/ai-and-the-future-of-cyber-competition/}.

\bibitem[Oesch et~al.(2024{\natexlab{a}})Oesch, Austria, Chaulagain, Weber,
  Watson, Dixson, and Sadovnik]{oesch2024path}
Sean Oesch, Phillipe Austria, Amul Chaulagain, Brian Weber, Cory Watson,
  Matthew Dixson, and Amir Sadovnik.
\newblock The path to autonomous cyber defense.
\newblock \emph{arXiv preprint arXiv:2404.10788}, 2024{\natexlab{a}}.

\bibitem[Oesch et~al.(2024{\natexlab{b}})Oesch, Chaulagain, Weber, Dixson,
  Sadovnik, Roberson, Watson, and Austria]{oesch2024towards}
Sean Oesch, Amul Chaulagain, Brian Weber, Matthew Dixson, Amir Sadovnik,
  Benjamin Roberson, Cory Watson, and Phillipe Austria.
\newblock Towards a high fidelity training environment for autonomous cyber
  defense agents.
\newblock In \emph{Proceedings of the 17th Cyber Security Experimentation and
  Test Workshop}, pages 91--99, 2024{\natexlab{b}}.

\bibitem[Pasquini et~al.(2024)Pasquini, Kornaropoulos, and
  Ateniese]{pasquini2024hacking}
Dario Pasquini, Evgenios~M Kornaropoulos, and Giuseppe Ateniese.
\newblock Hacking back the ai-hacker: Prompt injection as a defense against
  llm-driven cyberattacks.
\newblock \emph{arXiv preprint arXiv:2410.20911}, 2024.

\bibitem[Schmid(2018)]{schmid2018determinants}
Jon Schmid.
\newblock The determinants of military technology innovation and diffusion.
\newblock \emph{Scientific and Technological Flows Between the United States
  and China}, 2018.

\end{thebibliography}

\begin{IEEEbiography}{Sean Oesch}{\,}is a researcher at Oak Ridge National Laboratory, Oak Ridge, Tennessee. His current research interests include autonomous cyber defense, explainable AI for cyber applications, and AI for cyber defense. He is a Senior Member of IEEE. \vspace*{8pt}
\end{IEEEbiography}

\begin{IEEEbiography}{Jack Hutchins}{\,}is a researcher at Oak Ridge National Laboratory, Oak Ridge, Tennessee. His current research interests include AI security, robust AI, and edge AI. Jack recieved his M.S from the University of Tennessee, where he studied computer engineering with a focus on machine learning for electronic design automation. \vspace*{8pt}
\end{IEEEbiography}

\begin{IEEEbiography}{Phillipe Austria}{\,} is a researcher at Oak Ridge National Laboratory, Oak Ridge, Tennessee. His currently interest include data visualizations, machine learning and blockchain. Phillipe received his Ph.D. from the University of Nevada, Las Vegas, where he studied computer science with a focus on decentralized systems.\vspace*{8pt}
\end{IEEEbiography}

\begin{IEEEbiography}{Amul Chaulagain}{\,}is a software engineer working at Oak Ridge National Laboratory. He is currently involved in projects involving the utilization of autonomous systems and AI for cyber defense. Amul received his B.S. from Purdue University, where he studied Computer Science with a concentration on Software Engineering.
\end{IEEEbiography}

\vspace*{8pt}
\end{document}